\documentclass[a4paper,11pt]{article} \usepackage{graphicx}
\usepackage[american]{babel} 
\begin{document}

\date{}

\title{Generalized dimensions of Feigenbaum's attractor from
renormalization-group functional equations}

\author{Sergey P. Kuznetsov$^1$ and Andrew H. Osbaldestin$^2$}

\maketitle\begin{center} $^{1}$\textit{Institute of
Radio-Electronics, Russian Academy of Sciences,\\ Zelenaya 38,
Saratov 410019, Russian Federation}\\ \vspace{3mm}
$^{2}$\textit{Department of Mathematical Sciences Loughborough
University, \\Loughborough, LE11 3TU, UK}\end{center}

\begin{abstract}

A method is suggested for the computation of the generalized
dimensions of fractal attractors at the period-doubling
transition to chaos. The approach is based on an eigenvalue
problem formulated in terms of functional equations, with a
coefficient expressed in terms of Feigenbaum's universal fixed-point function.
The accuracy of the results is determined only by
precision of the representation of the universal function.

\end{abstract}


The multifractal or thermodynamic formalism is an important tool for
description of strange sets arising in dynamical systems in
different contexts. Its basic ideas have been clearly formulated
e.g. in the paper of Halsey et al. \cite{1}. Some of the examples
presented by these and other authors relate to the fractal
attractors that occur at the onset of chaos via period doubling
and quasiperiodicity~\cite{2}--\cite{7}. The multifractal analysis
reveals global scaling properties of these attractors, such as the
generalized dimensions and the $f(\alpha)$ spectra. They are of
principal interest because of their universality for systems of
different nature. Moreover, they allow a measurement in physical
experiments~\cite{7}.

One of the well-studied multifractal objects is the Feigenbaum
attractor, which occurs at the period-doubling transition to chaos
in unimodal one-dimensional maps with quadratic extremum and in a
wide class of more general nonlinear dissipative systems~\cite{8,2,9}.
Beside the original procedure of Halsey et al.
(namely the construction and analysis of the partition functions defined as
sums over some natural covering of the attractor), several other
approaches to the computation of the multifractal characteristics
have been developed. Bensimon et al.~\cite{3} used a method based on a
break up of a partition sum into two components with subsequent use
of the scaling property. Kov\'acs~\cite{4} suggested a procedure
of extracting the dimensions from the eigenvalue problem for the
Frobenius-Perron operator. Christiansen et al.~\cite{5} exploited
the idea of approximating the strange sets by periodic orbits and
expressed the desired quantities in terms of cycle expansions.
(To our knowledge, the calculation of the Hausdorff dimension of
the Feigenbaum attractor in Ref.~\cite{5} remains the most precise
to date.)

In some sense, the \textit{global} description of scaling properties in
the multifractal formalism seems opposite to the \textit{local}
description in terms of the Feigenbaum renormalization group
approach~\cite{8}. The latter is based on the solution of the functional
fixed-point equation and associated with scaling relations for the
evolution operators in a neighborhood of the extremum of the map
under consideration.

In this note we present a novel method for precise computation of
the multifractal characteristics: the problem may be presented in
terms of the Feigenbaum renormalization transformation applied to
some auxiliary function. The desired quantities, such as the
generalized dimensions and the $f(\alpha)$ spectrum, can be extracted
from an eigenvalue problem, formulated as a functional equation
involving Feigenbaum's universal fixed-point function.
An analogous approach was previously suggested in application to the
problem of the effect of noise onto the period-doubling transition
\cite{10,2}. That problem appears to be linked with one special
generalized dimension, as noted e.g. in Refs.~\cite{2, 11}, and
this circumstance obviously supports a possibility of the
generalization we undertake here. A similar idea was used in
Ref.~\cite{12} for a study of scaling regularities in the Fourier spectrum
and response function of the quadratic map at the period-doubling
transition to chaos.

Using the representation of the Feigenbaum function from Ref.~\cite{13}
we obtain the generalized dimensions with high
precision in excellent agreement with the previously known data.
Also we present accurate results for generalized dimensions of the
multifractal attractors at the onset of chaos in the unimodal maps
of degrees 4, 6, and 8, for which the universal functions $g(x)$
are available in the literature in the form of numerically found
polynomial expansions \cite{14}.

To estimate the multifractal characteristics for the Feigenbaum
attractor by the standard approach the generalized partition functions
$\Gamma_{k} (q,\,\tau ) = {\sum\limits_{i =1}^{2^k}p_i^q/l_i^\tau}$
are exploited. Here $q$ and $\tau$ are
some real parameters, $p_{i} = 2^{ - k}$,
$l_{i} = | x_i - x_{i + 2^k}|$, and the sequence $x_{i}$
results from iterations of the
unimodal map at the limit point of the period-doubling
accumulation, starting from the extremum point.
Obviously, $\Gamma _{k} (q,\,\tau ) = 2^{ - qk}S_{2^k}(\tau)$,
where $S_{2^k}(\tau) = {\sum\limits_{i = 1}^{2^k} {l_i^{-\tau}}}$.
For each given $q$ an appropriate value of $\tau = \tau (q)$ may
be found that ensures an asymptotic equality $\Gamma _{k + 1}
(q,\,\tau ) = \Gamma _{k} (q,\,\tau )$ as $k \to \infty $. Vice
versa, for a given~$\tau $ we can select a respective value of $q
= q(\tau)$. This relation of $q$ and $\tau $ is used then to
obtain the generalized dimensions and $f(\alpha)$ spectrum.

For large $k$ the lengths of the intervals $l_{i} $ are small, and
they may be estimated via the derivatives as
\begin{equation}
\label{eq1} l_{i} \cong |{d x_{i}}/ {d x_{1}}|l_1.
\end{equation}
In this approximation we can compute them together with the sums $S$ step by
step via simultaneous iterations of the relations
\begin{equation}
\label{eq2}
\begin{array}{l}
x_{i + 1} =
f(x_{i}),\\ l_{i + 1} = | f'(x_i)|l_i,\\ S_{i + 1} = S_i + l_{i +
1}^{ - \tau} \Psi (x_{i}).
\end{array} \end{equation}
We have introduced here an auxiliary function $\Psi (x)$ which at this moment
is supposed to be identically equal to 1.

By twofold application of Eqs. (\ref{eq2}) we obtain
\begin{equation}
\label{eq3} \begin{array}{l}
x_{i + 2} = f(f(x_{i})),\\ l_{i + 2} = |f'(f(x_i))f'(x_i)|l_i,\\
S_{i + 2} = S_i + l_{i + 2}^{-\tau} \left[|f'(f(x_i))|^{\tau} \Psi (x_i) +
\Psi (f(x_{i}))\right].
\end{array}
\end{equation}
Now we perform Feigenbaum's scale change $x \mapsto x/\alpha$, $l
\mapsto l/|\alpha|$
($\alpha$ is the Feigenbaum constant) and arrive at the equations,
which are of the same form as (\ref{eq2}), but with new functions
$f_{\mathrm{new}}(x) = \alpha f(f(x/\alpha))$,
$\Psi_{\mathrm{new}}(x) =L_{f,\tau}\Psi(x)$ where $L_{f,\tau}$ is the
linear operator
\begin{equation}
L_{f,\tau}:\Psi(x)\mapsto |\alpha|^{\tau}
\left[|f'(f(x/\alpha))|^{\tau}\Psi(x/\alpha)+\Psi(f(x/\alpha))
\right].
\end{equation}

This transformation may be repeated again and again.
Asymptotically, $f(x)$ converges to the
fixed-point function satisfying the Feigenbaum-Cvitanovi\'c
equation $g(x) = \alpha g(g(x/\alpha))$, and the
sequence $\Psi _{k}(x)$ will follow the recursion
$\Psi_{k+1}(x)=L_{g,\tau}\Psi_k(x)$.

Thus, as $k \to \infty $, the function $\Psi_k$ tends to the
eigenfunction associated with the largest eigenvalue of the
linear operator $L_{g,\tau}$, the eigenproblem being
\begin{equation}
\label{eq6}
\nu(\tau) \Psi(x)= |\alpha|^{\tau}
\left[|g'(g(x/\alpha))|^{\tau}\Psi(x/\alpha)+\Psi(g(x/\alpha))\right].
\end{equation}
(A particular case of this equation for $\tau = 2$
appears in the theory of effect of noise onto the period-doubling
transition. A possibility of computation of the noise scaling
constant via sums of the derivatives over the Feigenbaum attractor
was noted e.g. in Refs. \cite{15,11}.)

>From the construction, we see that the eigenvalue $\nu(\tau)$
indicates a rate of growth or decrease of the sums $S$:
\begin{equation}
\label{eq7}
S_{2^{k}} (\tau ) \propto \nu(\tau)^{k}.
\end{equation}
To have $\Gamma _{k} \to {\rm const}$ as
$k \to \infty$, we must have
\begin{equation}
\label{eq7a}
\nu (\tau)
= 2^{q}{\rm ,} \,\,\,{\rm or}\,\,\, q = \log _{2} \nu (\tau ).
\end{equation}
Then, in accordance with the
multifractal formalism, we can obtain the generalized dimensions
\begin{equation}
\label{eq8}
D_{q} = \frac{\tau}{q-1}\,,
\end{equation}
and the $f(\alpha)$ spectrum as an implicitly
defined relation between the variables
\begin{equation}
\label{eq8a}
\alpha = \frac {d \tau}{dq}
\,\,\,{\rm and} \,\,\,f = q\frac{d\tau}{dq} - \tau.
\end{equation}

Although our argumentation starts from the approximate relation
(\ref{eq1}), we believe that the final Eq.~(\ref{eq6}) is exact.
The data from the numerical computations presented below strongly
supports this conjecture: the generalized dimensions are in
excellent agreement with the best known numerical results, up to
all reliable digits. Apparently, in the asymptotics of $k \to
\infty $ the approximate nature of (\ref{eq1}) becomes
inessential. One might hope that a rigorous proof can be found.

We have performed numerical solutions of the eigenvalue problem
(\ref{eq6}) for the classic Feigenbaum attractor of the quadratic
map and for unimodal maps of even integer degrees $d$ = 4, 6,
and 8. In principle, the achievable precision of the results is
determined only by accuracy of the approximation of the
universal functions.

With the known polynomial approximation of $g(x)$ and value of
the scaling constant~$\alpha$ we have numerically performed the functional
transformation defined by the right-hand side of Eq.~(\ref{eq6}).
The unknown function $\Psi (x)$ is represented
by a table of its values at the nodes of a one-dimension grid on
the interval $[0,1]$ and by an interpolation scheme between the
nodes. In actual computations it was convenient to use a grid of
constant step along the axis of variable $y = |x|^d$ and a
fourth-order interpolation in terms of $y$. Given the input
table for $\Psi (x)$ the program yields an analogous table as
output.

Suppose we fix $\tau$ and wish to estimate $q$. We define an
initial condition as $\Psi (x) \equiv 1$, perform the functional
transformation, and normalize the resulting function as $\Psi
^0(x) = \Psi (x)/\Psi (0)$. The new function is taken as the
initial condition and so on. This operation is repeated many
times, until the form of the function $\Psi (x)$ stabilizes. Then,
the value of $\Psi (0)$ before the normalization is taken to be
$\nu(\tau)$, and we finally set $q(\tau ) = \log _2 \nu(\tau)$.

To find $\tau$ for a given $q$ we use the above procedure together
with a simple iteration scheme for the numerical solution of the
algebraic equation $q(\tau )=q$. We may then
calculate $D_q = \tau/(q-1)$ at $q \ne 1$. In particular, $D_{0}$
is the Hausdorff dimension, and $D_{2}$ is the correlation
dimension.

To obtain the information dimension $D_{1} $ it is necessary to
determine the limit as $q \to 1$, that is, as $\tau \to 0$.
Formally, this follows from L'H\^opital's rule: $D_{1} = \lim\limits_{q
\to 1}\frac{\tau (q)}{q - 1} = \left( \frac{d\tau}{dq} \right)_{q
= 1} = \left(\frac{dq}{d\tau}\right)_{\tau = 0}^{ - 1}.$ To
compute this without loss of accuracy we use the following
algorithm. For $\tau \ll 1$ let us write $\Psi _{k} (x) = 2^{k}|
\alpha|^{k\tau} [1 + \tau h_k(x)]$ and substitute this expression
into Eq.~(\ref{eq5}). To first order we have
\begin{equation}
\label{eq9}
h_{k + 1} (x) = \textstyle{1\over
2}\left[ h_k (x/\alpha)+ h_k (g(x/\alpha))\right] +
\textstyle{1\over 2}\ln |g'(g(x/\alpha))|\,.
\end{equation}
Numerically, representing $h_{k}(x)$ by a table of its values and
performing a large number of steps of the transformation one can
observe that $h_{k + 1} (x) - h_{k} (x)\to\theta = {\rm const}$
as $k\to\infty$. This implies that $\Psi _{k}
\propto |\alpha|^{k\tau} 2^{k}e^{k\theta \tau}$. On the other hand,
$\Psi_k \propto 2^{kq(\tau)} \cong 2^{k(q + \tau dq/d\tau)}
=2^{k(1+\tau/D_1)}$. Hence,
\begin{equation}
\label{eq10}
D_{1} = \frac{\ln 2}{\ln|\alpha|+\theta}.
\end{equation}

For quadratic maps we have performed the computations based on a
polynomial representation of $g(x)$ with coefficients taken from
the paper of Lanford~\cite{13}. His data are of very high precision, but in our
calculations the accuracy is limited due to a use of standard
double-precision arithmetic. As a result, we get not more than 14 true
digits in the generalized dimensions. These data are presented in
the first column of Table~1. Note excellent agreement of the 
Hausdorff dimension (up to
the last decimal digit!) with the result of Christiansen et al.~\cite{5}.
Other dimensions for the Feigenbaum attractor were presented
by Kov\'acs~\cite{4}, and they coincide with our results up to the
10-th digit, the accuracy achieved in that work.

In the remaining three columns of the Table~1 we present
results for the generalized dimensions of the
multifractal attractors at the onset of chaos in unimodal maps
of degrees 4, 6, and 8 obtained using the universal functions
given e.g. in Ref.\cite{14}.

As an alternative to the traditional definition of the generalized
dimensions $D_{q}$ one might consider a family of dimensions
enumerated by the index $\tau$. Let us designate them as $D^{\tau}$:
$D^{\tau} = D_{q(\tau)} = \tau /(q(\tau)-1)$. As mentioned, for
$\tau=2$ the equation (\ref{eq6}) is of a form studied in the
theory of noise effect onto the period-doubling transition
\cite{10}; the noise scaling constant is defined as $\gamma =
\sqrt {\nu(2)} $. Hence, the dimension $D^2$ is linked with the
effect of noise. The scaling factor $\gamma$ is
expressed via the dimension $D^2$ as $\gamma = 2^{1/D^2+1/2}$. In
Table~2 we present high-precision data for values of $q(2)$,
dimensions $D^2$, and noise scaling factors obtained from the
numerical solution of the eigenproblem (\ref{eq6}) for maps of
degree 2, 4, 6, and 8. The data for the factors $\gamma$ improve
the previously known results~\cite{16}. (Observe that value of $q$
depends on the degree, so one cannot speak on a definite
dimension from the family $D_q$ associated with the noise effect!)

For the problem of unimodal maps, our method of calculation of the
generalized dimensions
does not have obvious computational advantages over those of
Refs.~\cite{3,4,5}, but it does represent the problem in a new light and
indicates novel links between global and local descriptions of the
scaling regularities. An analogous use of renormalization-group
equations will be feasible in the calculation of
multifractal properties in many other situations at the onset of
chaos, e.g. in bimodal
one-dimensional maps~\cite{17}, asymmetric one-dimensional maps~\cite{18},
two-dimensional period-doubling maps~\cite{19},
quasiperiodically forced maps~\cite{20}, and complex analytic maps~\cite{21}.
Such an approach will be useful especially for situations
where computations based on the traditional partition-function approach are
difficult.

\vspace{3mm} \textbf{Acknowledgements}\vspace{3mm}

The authors acknowledge support from the London Mathematical Society.

\newpage

\begin{table}[!ht] \caption{Generalized dimensions for fractal
attractors at the onset of chaos in unimodal maps of degree $d$}
\vspace{3mm} \begin{tabular} {|c|c|c|c|c|} \hline

& $d = 2$ & $d = 4$ & $d = 6$ & $d = 8$\\

\hline

$D_5$ & 0.45392270234470 & 0.407695571 & 0.373232166 & 0.351475400\\
\hline

$D_4$ & 0.46615155691823 & 0.426832904 & 0.392400635 & 0.370142909\\
\hline

$D_3$ & 0.48077684940009 & 0.454569793 & 0.421638052 & 0.399231039\\
\hline

$D_2$ & 0.49783645928917 & 0.495316676 & 0.468035066 & 0.447019466\\
\hline

$D_1$ & 0.51709757255124 & 0.555181822 & 0.544847134 & 0.531111008\\
\hline

$D_0$ & 0.53804514358055 & 0.642575065 & 0.683433256 & 0.707102082\\
\hline

$D_{-1}$ &0.55991291016494 & 0.763919555 & 0.946229117 &
1.146118382\\ \hline

$D_{-2}$ & 0.58173600034603 & 0.894257449 & 1.205507002 &
1.510079742\\ \hline

$D_{-3}$ & 0.60247817187829 & 0.992066238 & 1.354808070 &
1.698747772\\ \hline

$D_{-4}$ & 0.62126594260209 & 1.056616863 & 1.445090859 &
1.811996998\\ \hline

$D_{-5}$ & 0.63760518368338 & 1.100453275 & 1.505301852 &
1.887496869\\ \hline

\end{tabular} \end{table}

\begin{table}[!ht] \caption{Generalized dimensions $D^2$
and noise scaling factors for unimodal maps of degree $d$}
\vspace{3mm}

\begin{tabular}{|c|c|c|c|c|}
\hline & $d=2$ & $d=4$ & $d=6$ & $d=8$\\
\hline $q(2)$& 5.45324245756108 & 6.086657808 & 6.654767241 &
7.070578662\\
\hline $D^2=D_{q(2)}$ & 0.44911096107158 & 0.393185481 &
0.353683877 & 0.329457884 \\
\hline $\gamma$ & 6.61903651081803 &
8.243910853 & 10.037886410 & 11.59386214 \\ \hline \end{tabular}
\end{table}


\begin{thebibliography}{99}

\bibitem{1} T.S.Halsey, M.H.Jensen, L.P.Kadanoff, I.Procaccia,
and B.I.Shraiman. Fractal measures and their singularities: the
characterization of strange sets. Phys. Rev. A, \textbf{33},
1141--1151 (1986).

\bibitem{2} E.B.Vul, Y.G.Sinai, and K.M.Khanin. Feigenbaum
universality and thermodynamic formalism. Russ. Math. Surv.,
\textbf{39}(3), 1-40 (1984). [Russian original: Usp. Mat. Nauk,
39(3), 3--37 (1984).]

\bibitem{3} D.Bensimon, M.H.Jensen, and L.P.Kadanoff.
Renormalization-group analysis of the period-doubling attractor.
Phys. Rev. A, \textbf{33}, 3622--3624 (1986).

\bibitem{4} Z.Kov\'acs. Universal $f(\alpha)$ spectrum as an
eigenvalue. J. Phys. A: Math. Gen., \textbf{22}, 5161--5165 (1989).

\bibitem{5} F.Christiansen, P.Cvitanovi\'c, and H.H.Rugh. The
spectrum of the period-doubling operator in terms of cycles. J.
Phys. A: Math. Gen., \textbf{23}, L713--L717 (1990).

\bibitem{6} A.H.Osbaldestin. Siegel disk singularity spectra. J.
Phys. A: Math. And Gen., \textbf{25}, 1169--1175 (1992);
A.D.Burbanks, A.H.Osbaldestin, and A.Stirnemann. Fractal dimension of
Siegel disc boundaries. European Phys. J., \textbf{B4}, 263--265
(1998).

\bibitem{7} J.A.Glazier, M.H.Jensen, A.Libchaber, and J.Stavans.
Structure of Arnold tongues and the $f(\alpha)$ spectrum for
period doubling -- Experimental results. Phys. Rev. A, \textbf{34},
1621--1624 (1986); Z.Su, R.W.Rollins, and E.R.Hunt. Measurements of
$f(\alpha)$ spectrum in driven diode resonator systems. Phys. Rev.
A, \textbf{36}, 3515--3517 (1987); J.A.Glazier, G.Gunaratne, and A.Libchaber.
$f(\alpha)$ curves -- experimental results. Phys. Rev. A, \textbf{37},
523-530 (1988); R.E.Ecke, R.Mainieri, and T.S.Sullivan. Universality
in quasi-periodic Rayleigh-Benard convection. Phys. Rev. A,
\textbf{44}, 8103--8118 (1991).

\bibitem{8} M.J.Feigenbaum. Quantitative universality for a
class of nonlinear transformations. J. Stat. Phys., \textbf{19},
25--52 (1978); M.J.Feigenbaum. The universal metric
properties of nonlinear transformations. J. Stat. Phys.
\textbf{21}, 669--706, (1979); M.J.Feigenbaum. Universal
behavior in nonlinear systems. Physica \textbf{D7},
16--39 (1983).

\bibitem{9} P.Cvitanovi\'c, ed. Universality in Chaos. Adam
Hilger, 2nd Edition (1989), 631p.

\bibitem{10} J.P.Crutchfield, M.Nauenberg, and J.Rudnik. Scaling for
external noise at the onset of chaos. Phys. Rev. Lett., \textbf{46},
933--935 (1981); B.Shraiman, C.E.Wayne, and P.C.Martin. Scaling theory
for noisy period-doubling transitions to chaos. Phys. Rev. Lett.,
\textbf{46}, 935--939 (1981).

\bibitem{11} A.Hamm and R.Graham. Scaling for small random
perturbations of golden critical circle maps. Phys. Rev. A,
\textbf{46}, 6323--6333 (1992).

\bibitem{12} S.P.Kuznetsov and A.S.Pikovsky. Renormalization group
for the response function and spectrum of the period-doubling
system. Phys. Lett., \textbf{A140}, 166--172 (1989).

\bibitem{13} O.E.Lanford III. A computer assisted proof of the
Feigenbaum conjectures. Bull. Amer. Math. Soc., \textbf{6},
427--434 (1982).

\bibitem{14} A.P.Kuznetsov, S.P.Kuznetsov, and I.R.Sataev.
Three-parameter scaling for one-dimensional maps. Phys. Lett.,
\textbf{A189}, p.367--373 (1994).

\bibitem{15} T.Kai. Lyapunov Number for a noisy $2^n$ cycle. J. Stat. Phys.,
\textbf{29}, 329--343 (1982).

\bibitem{16} P.Collet and A.Lesne. Renormalization-group analysis
of some dynamical systems with noise. J. Stat. Phys., \textbf{57},
967--992 (1989); A.P.Kuznetsov and J.V.Kapustina. Properties of
scaling at the transition to chaos in model maps with noise.
Izvestija VUZov -- Prikladnaya Nelineinaya Dinamika,
\textbf{8}(\ref{eq6}), 78--87 (2000).

\bibitem{17} R.S.MacKay and J.B.J. van Zeijts. Period doubling for
bimodal maps: A horseshoe for a renormalization operator.
Nonlinearity, \textbf{1}, 253--277 (1998).

\bibitem{18} B.D.Mestel and A.H.Osbaldestin. Asymptotics of scaling
parameters for period-doubling in unimodal maps with asymmetric
critical points. J. Math. Phys., \textbf{41}, 4732--4746 (2000).

\bibitem{19} A.P.Kuznetsov, S.P.Kuznetsov, and I.R.Sataev. A variety
of period-doubling universality classes in multi-parameter
analysis of transition to chaos. Physica \textbf{D109}, 91--112
(1997).

\bibitem{20} S.P.Kuznetsov, U.Feudel and A.S.Pikovsky.
Renormalization group for scaling at the torus-doubling terminal
point. Phys. Rev. E, \textbf{57}, 1585--1590 (1998); S.P.Kuznetsov,
E.Neumann, A.Pikovsky, and I.R.Sataev. Critical point of
tori-collision in quasiperiodically forced systems. Phys. Rev. E,
\textbf{62}, 1995--2007 (2000).

\bibitem{21} A.I.Golberg, Y.G.Sinai, and K.M.Khanin. Universal
properties of the period-tripling sequence. Usp. Mat. Nauk,
\textbf{38}, 159--160 (1983); P.Cvitanovi\'c, J.Myrheim. Complex
universality. Comm. Math. Phys., \textbf{121}, 225--254 (1989);
B.D.Mestel and A.H.Osbaldestin. Renormalization in implicit complex
maps. Physica \textbf{D39}, 149--162 (1989)


\end{thebibliography}
\end{document}